\documentclass[aps,prl,preprint,tightenlines,superscriptaddress,showpacs,byrevtex,amsmath,amssymb]{revtex4}
\usepackage{epsfig}
\usepackage{dcolumn}  

\graphicspath{{ps}}

\begin{document}

\preprint{\vbox{\hbox{\hfil Belle Prerpint 2005-15}
                \hbox{\hfil KEK   Preprint 2005-7}}}

\title{
 \quad\\[0.5cm]
Studies of CP violation in $B \rightarrow J/\psi K^{*}$ decays}

\affiliation{Budker Institute of Nuclear Physics, Novosibirsk}
\affiliation{Chiba University, Chiba}
\affiliation{Chonnam National University, Kwangju}
\affiliation{University of Cincinnati, Cincinnati, Ohio 45221}
\affiliation{Gyeongsang National University, Chinju}
\affiliation{University of Hawaii, Honolulu, Hawaii 96822}
\affiliation{High Energy Accelerator Research Organization (KEK), Tsukuba}
\affiliation{Institute of High Energy Physics, Chinese Academy of Sciences, Beijing}
\affiliation{Institute of High Energy Physics, Vienna}
\affiliation{Institute for Theoretical and Experimental Physics, Moscow}
\affiliation{J. Stefan Institute, Ljubljana}
\affiliation{Kanagawa University, Yokohama}
\affiliation{Korea University, Seoul}
\affiliation{Kyungpook National University, Taegu}
\affiliation{Swiss Federal Institute of Technology of Lausanne, EPFL, Lausanne}
\affiliation{University of Ljubljana, Ljubljana}
\affiliation{University of Maribor, Maribor}
\affiliation{University of Melbourne, Victoria}
\affiliation{Nagoya University, Nagoya}
\affiliation{Nara Women's University, Nara}
\affiliation{National Central University, Chung-li}
\affiliation{National United University, Miao Li}
\affiliation{Department of Physics, National Taiwan University, Taipei}
\affiliation{H. Niewodniczanski Institute of Nuclear Physics, Krakow}
\affiliation{Nippon Dental University, Niigata}
\affiliation{Niigata University, Niigata}
\affiliation{Nova Gorica Polytechnic, Nova Gorica}
\affiliation{Osaka City University, Osaka}
\affiliation{Osaka University, Osaka}
\affiliation{Panjab University, Chandigarh}
\affiliation{Peking University, Beijing}
\affiliation{Princeton University, Princeton, New Jersey 08544}
\affiliation{Saga University, Saga}
\affiliation{University of Science and Technology of China, Hefei}
\affiliation{Seoul National University, Seoul}
\affiliation{Sungkyunkwan University, Suwon}
\affiliation{University of Sydney, Sydney NSW}
\affiliation{Tata Institute of Fundamental Research, Bombay}
\affiliation{Toho University, Funabashi}
\affiliation{Tohoku Gakuin University, Tagajo}
\affiliation{Tohoku University, Sendai}
\affiliation{Department of Physics, University of Tokyo, Tokyo}
\affiliation{Tokyo Institute of Technology, Tokyo}
\affiliation{Tokyo Metropolitan University, Tokyo}
\affiliation{Tokyo University of Agriculture and Technology, Tokyo}
\affiliation{University of Tsukuba, Tsukuba}
\affiliation{Virginia Polytechnic Institute and State University, Blacksburg, Virginia 24061}
\affiliation{Yonsei University, Seoul}
   \author{R.~Itoh}\affiliation{High Energy Accelerator Research Organization (KEK), Tsukuba} 
   \author{Y.~Onuki}\affiliation{Niigata University, Niigata} 
   \author{K.~Abe}\affiliation{High Energy Accelerator Research Organization (KEK), Tsukuba} 
   \author{K.~Abe}\affiliation{Tohoku Gakuin University, Tagajo} 
   \author{I.~Adachi}\affiliation{High Energy Accelerator Research Organization (KEK), Tsukuba} 
   \author{H.~Aihara}\affiliation{Department of Physics, University of Tokyo, Tokyo} 
   \author{Y.~Asano}\affiliation{University of Tsukuba, Tsukuba} 
   \author{T.~Aushev}\affiliation{Institute for Theoretical and Experimental Physics, Moscow} 
   \author{A.~M.~Bakich}\affiliation{University of Sydney, Sydney NSW} 
   \author{Y.~Ban}\affiliation{Peking University, Beijing} 
% ...
\collaboration{The Belle Collaboration}
\noaffiliation



\begin{abstract}
CP violation in $B\rightarrow J/\psi K^*$ decays is studied using
an angular analysis in a data sample of 253 fb$^{-1}$ 
recorded with the Belle detector at the KEKB $e^+e^-$ collider.
The flavor separated measurements of the decay
amplitudes indicate no evidence for direct CP violation. 
T-odd CP violation is studied using
the asymmetries in triple product correlations, and the results are
consistent with the Standard Model null predictions. 
The time-dependent angular analysis gives the 
following values of CP-violating parameters:
$\sin 2\phi_1 = 0.24 \pm 0.31 \pm 0.05$ and 
$\cos 2\phi_1 = 0.56 \pm 0.79 \pm 0.11$.
\end{abstract}


\maketitle

{\renewcommand{\thefootnote}{\fnsymbol{footnote}}}
\setcounter{footnote}{0}


An angular analysis of $B$ meson decay to two vector mesons is a
sensitive probe of new physics. 
There are three classes of parameters 
obtained through the angular analysis. The first is the 
measurement of the decay amplitudes of the three helicity states. 
These can be obtained by the time-integrated
angular analysis of flavor specific decays. 
The comparison of the amplitudes between flavor separated samples
probes direct CP violation. 
The second is the triple product correlation, which can be 
extracted from the measured decay amplitudes.
This quantity is sensitive to T-odd CP violation. 
The third class is comprised of the
CP parameters ($\sin 2\phi_1$ and $\cos 2\phi_1$) 
that are measured
through a time-dependent angular analysis. In particular, the
measurement of $\cos 2\phi_1$, which appears in the time-dependent
interference terms, is important
both to solve the two-fold ambiguity in $2\phi_1$ and to test the
consistency of this determination with the more precise value from
other $b \to c\bar{c}s$ decays.
In this paper, we report the measurements of these
parameters for $B \rightarrow J/\psi K^{*}$ decays. 

\begin{figure}
\centerline{\mbox{\psfig{figure=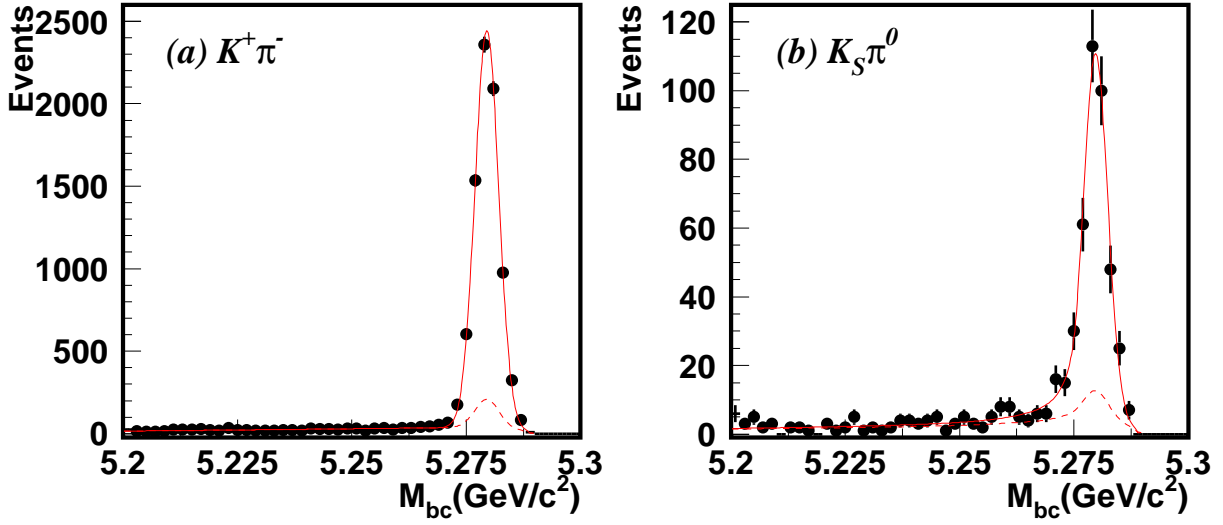,width=16cm}}}
\caption{ $M_{\rm bc}$ distributions for the candidates of 
(a) $B^0\rightarrow J/\psi K^{*0}(K^+\pi^-)$, and
(b) $B^0\rightarrow J/\psi K^{*0}(K_S^0\pi^0)$.
The solid line shows
the projection of the two-dimensional fit to $M_{\rm bc}$ and $\Delta E$ 
while the dashed line is the 
estimated background contamination.}
\label{dEMbcplot}
\end{figure}

The data sample used in this analysis corresponds to an integrated
luminosity of 253 fb$^{-1}$ recorded with the Belle detector~\cite{Belle}
at the KEKB electron-positron collider~\cite{KEKB}.
Four decay modes are reconstructed:
$B^0\rightarrow J/\psi  K^{*0}; K^{*0}\rightarrow K^+\pi^-$ and
$K_S^0\pi^0$, and $B^+ \rightarrow J/\psi K^{*+};
K^{*+} \rightarrow K^+\pi^0$ and $K_S^0\pi^+$. The charge
conjugate modes are included everywhere unless otherwise specified.
The reconstruction is done using the criteria described in
Ref.~\cite{angleana-29}.
A $J/\psi$ candidate is reconstructed from two oppositely charged
tracks identified as leptons.
A $K^{*}$ candidate is selected if the absolute difference between
the invariant mass of
an identified $K \pi$ pair and the nominal $K^*(892)$ mass
is less than $75~{\rm MeV}/c^2$.  
Candidate $B$ mesons are reconstructed by combining a
$J/\psi$ candidate with a $K^*$ candidate
and examining two
quantities in the center-of-mass of the $\Upsilon$(4S): the
beam-constrained mass calculated using the beam energy 
in place of the reconstructed energy ($M_{\rm bc}$),
and the energy difference between the $B$
candidate and the beam energy ($\Delta E$).
$M_{\rm bc}$ is required to be in the range 5.27-5.29$~{\rm GeV}/c^2$.
For the modes with a charged (neutral) pion,
the energy difference must satisfy $|\Delta E| < 30\,{\rm MeV}$
($-50\,{\rm MeV}<\Delta E< 30\,{\rm MeV}$).
To eliminate slow $\pi^0$ backgrounds,
the angle $\theta_{K^*}$ of the kaon with respect
to the opposite of $B$ direction
in the  $K^*$ rest frame is required
to satisfy ${\rm cos}\theta_{K^*}<0.8$. 
When an event contains more than one candidate passing the above
requirements, 
the best combination is selected based on a
$\chi^2$ calculated using $M_{\rm bc}$ and $\Delta E$.
Figure ~\ref{dEMbcplot} shows the $M_{\rm bc}$ distributions
for the candidates of two neutral $B$ decay modes within the $\Delta E$ signal
window. 
After all selections are applied, the remaining numbers of 
events in the signal region of the $M_{\rm bc}$--$\Delta E$ plane 
are 8194 for $J/\psi
K^{*0}(K^+\pi^-)$, 363 for $J/\psi K^{*0}(K_S^0\pi^0)$, 2222 for $J/\psi
K^{*+}(K^+\pi^0)$ and 2168 for $J/\psi K^{*+}(K_S^0\pi^+)$.


The angular distribution of the products
of $B\rightarrow J/\psi K^*$ decays is described using
three angles in the transversity basis~\cite{transversity}.
The direction of motion of the $J/\psi$ in the rest frame of the $B$
candidate is defined to be the $x$-axis. The $y$-axis is chosen along
the direction of the projection of the $K$ momentum into the plane
perpendicular to the $x$-axis in the $B$ rest frame. 
The $z$-axis is then perpendicular to the $x$-$y$ plane
according to the right-hand rule.
The angle between the positive lepton ($l^+$) and the
$z$-axis 
in the $J/\psi$ rest frame is defined as $\theta_{tr}$.
The angle between the $x$-axis and the projection
of the $l^+$ momentum
onto the $x$-$y$ plane is defined as $\phi_{tr}$ in the same frame. 
The angle $\theta_{K^*}$ is defined earlier. 

The distribution of the 
angles as a function of the decay time difference between $B$ and
anti-$B$ mesons ($\Delta t$)
is described as follows~\cite{Yamamoto}:
\begin{gather}
\frac{1}{\Gamma}\frac{d^4\Gamma(\theta_{tr},\phi_{tr},\theta_{K^*},\Delta t)}
{d\cos\theta_{tr} d\phi_{tr} d\cos\theta_{K^*} d\Delta t}
 =  \frac{9}{32\pi}  
	\frac{e^{-|\Delta t|/\tau_{B}}}{2\tau_{B}} \sum_{i=1}^{6} 
g_i (\theta_{tr},\phi_{tr},\theta_{K^*}) a_i(\Delta t)
\label{eqn:tadf}
\end{gather}
where the angular terms $g_i$ are defined as
\begin{eqnarray*}
g_1 & = & 2\cos^2\theta_{K^*}(1-\sin^2\theta_{tr}\cos^2\phi_{tr}) \\
g_2 & = & \sin^2\theta_{K^*}(1-\sin^2\theta_{tr}\sin^2\phi_{tr}) \\
g_3 & = & \sin^2\theta_{K^*}\sin^2\theta_{tr} \\
g_4 & = & -(1/\sqrt{2})\sin 2\theta_{K^*}\sin^2\theta_{tr}\sin2\phi_{tr} \\
g_5 & = & \sin^2\theta_{K^*}\sin2\theta_{tr}\sin\phi_{tr} \\
g_6 & = & (1/\sqrt{2})\sin
2\theta_{K^*}\sin2\theta_{tr}\cos\phi_{tr} ,
\end{eqnarray*}
and the amplitude terms $a_i$ as 
\begin{eqnarray*}
a_1 & = & |A_0|^2(1+\eta\sin2\phi_1 \sin\Delta m\Delta t), \\
a_2 & = & |A_{\parallel}|^2(1+\eta\sin2\phi_1 \sin\Delta m\Delta t), \\
a_3 & = & |A_{\perp}|^2(1-\eta\sin2\phi_1 \sin\Delta m\Delta t), \\
a_4 & = & {\rm Re}(A_{\parallel}^*A_0)(1+\eta \sin2\phi_1 \sin\Delta
m\Delta t), \\
a_5 & = & \eta {\rm Im}(A_{\parallel}^*A_{\perp})\cos \Delta m\Delta t \\
& & -
\eta {\rm Re}(A_{\parallel}^*A_{\perp})\cos 2\phi_1 \sin \Delta m \Delta t, \\
a_6 & = & \eta {\rm Im}(A_0^*A_{\perp})\cos \Delta m\Delta t  \\
& & - \eta {\rm Re}(A_0^*A_{\perp})\cos 2\phi_1 \sin \Delta m \Delta t.
\end{eqnarray*}
Here, $A_0$, $A_{\parallel}$ and $A_{\perp}$ are the complex decay
amplitudes of the three helicity states in the transversity basis, and
$\eta = +1~(-1)$ for $B^0$ or $B^+$ ($\overline{B}^0$ or $B^-$). 
$\Delta m$ is the $B^0-\overline{B}^0$ mixing parameter, which is zero 
for charged $B$ meson decays, and
$\tau_{B}$ is the lifetime of a $B$ meson.
$\Gamma$ is the decay rate to each final state.
Two CP violation parameters appear in the formula,
{\it viz}, $\sin2\phi_1$ and $\cos2\phi_1$. They can take non-zero
values for the decay into CP eigenstate 
$B^0 \rightarrow J/\psi K^{*0}(K_S^0\pi^0)$ that occurs through the 
same quark level diagram as that of the golden mode for $\sin 2\phi_1$ 
measurement, $B^0\rightarrow J/\psi K_S^0$. 


The determination of the decay amplitudes and CP parameters are
performed using an unbinned maximum likelihood method, taking into
account the detection efficiency and backgrounds.
The probability density function (PDF) for an event is defined as
\begin{eqnarray}
{\mathcal P} & = & 
f_{sig}(M_{\rm bc},\Delta E)
\epsilon(\theta_{tr},\phi_{tr},\theta_{K^*})
   \times \frac{1}{\Gamma}
\frac{d^4\Gamma(\theta_{tr},\phi_{tr},\theta_{K^*},\Delta t)}
{d\cos\theta_{tr} d\phi_{tr} d\cos\theta_{K^*} d\Delta t}  \nonumber \\
& & + \frac{e^{-|\Delta t|/\tau_{B}}}{2\tau_{B}} \left\{
  \sum_i f_{cf}^i(M_{\rm bc},\Delta E)
{\mathcal A}_{cf}(\theta_{tr},\phi_{tr},\theta_{K^*}) \right.  \nonumber \\
& & ~~~~~~~~~~~~~~~~ \left. + f_{nr}(M_{\rm bc},\Delta E) 
{\mathcal A}_{nr}(\theta_{tr},\phi_{tr},\theta_{K^*}) \rule[-4mm]{0mm}{6mm}\right\}
\nonumber \\
& & + \delta(\Delta t) f_{cb}(M_{\rm bc},\Delta E) 
{\mathcal A}_{cb}(\theta_{tr},\phi_{tr},\theta_{K^*}), 
\label{eqn:timepdf}
\end{eqnarray}
where $f_{sig}$, $f_{cf}$, $f_{nr}$ and $f_{cb}$ are the respective
fractions of 
signal, cross-feeds, non-resonant production, and combinatorial
background components as functions of $\Delta E$ and $M_{bc}$, 
while
${\mathcal A}_{cf}$, ${\mathcal A}_{nr}$ and ${\mathcal A}_{cb}$ are
the corresponding angular shape functions, and
$\epsilon$ is the detection efficiency function.

Three separate background sources are considered in the fit:
cross-feeds that are contaminations from other $K^*$ subdecays, 
non-resonant production of $K\pi$ including contaminations
from higher resonance tails, and combinatorial
background. 
The fraction and angular shape of cross-feeds are 
estimated from Monte Carlo.
The fraction of non-resonant 
production is estimated to be 6.8\% from a fit to the invariant mass 
distribution of $K\pi$ pairs
in an alternate data sample that is assembled without the 
$K^{*}$ selection criteria.
The fit is performed with 
two Breit-Wigner functions for the $K^*(892)$ and $K^*_2(1430)$
resonances and a threshold function describing the non-resonant
production, taking into account the contaminations of 
cross-feeds and combinatorial background.
The phase space factor includes
a $J/\psi$ recoil correction. The angular shape is obtained from samples in
the mass region $1.0\,{\rm GeV}/c^2<M(K\pi)<1.3\,{\rm GeV}/c^2$. 

The fractions of signal and  
combinatorial background are estimated from 
a fit to the samples in the region of 
$5.2\,{\rm GeV}/c^2<M_{\rm bc}< 5.29\,{\rm GeV}/c^2$
and $-0.1\,{\rm GeV}< \Delta E < 0.1\,{\rm GeV}$.
The angular shape of combinatorial background is obtained from the
sub-sample with $5.2\,{\rm GeV}/c^2 < M_{\rm bc} < 5.26\,{\rm GeV}/c^2$ 
used in the fit above.
The fractions of both signal and backgrounds are parameterized 
as two-dimensional functions of $M_{\rm bc}$ and $\Delta E$.
The background angular shapes are
parameterized for each of three angles separately. 
For each signal mode, the detection efficiency is 
parameterized as a three-dimensional function whose parameters
are obtained by a fit to a high-statistics
Monte Carlo sample. The function is almost flat 
except in the region $\cos \theta_{K^*} \sim 1$, where the pion is
slow so that the efficiency is reduced.


The decay amplitudes are determined by fitting the time-integrated
angular distribution to three measured angles.
Eqs.~\ref{eqn:tadf} and \ref{eqn:timepdf} are integrated over
$\Delta t$,
where terms with $\frac{e^{-|\Delta
t|/\tau_{B}}}{2\tau_{B}}$ and $\cos \Delta m \Delta t$ become
unity while terms with 
$\sin \Delta m \Delta t$ become 0.
The value of $\eta$ is determined from the charge of the 
kaon for $K^*$ decays with a $K^+$ or of the pion with a $K_S^0$.
The decay mode $B^0\rightarrow
J/\psi K^{*0}(K_S^0 \pi^0)$ is not used, since $\eta$ cannot be
determined by this prescription.
In the fit, the imaginary part of $A_0$ is defined to be zero
since the overall phase of the decay amplitudes is arbitrary.
The values of the other five parameters, 
$|A_0|^2$, $|A_{\parallel}|^2$, $|A_{\perp}|^2$,
$\arg(A_{\parallel})$ and $\arg(A_{\perp})$,
are determined in the fit.
There is a two fold ambiguity in $\arg(A_{\parallel})$ and
$\arg(A_{\perp})$~\cite{Mahiko}. We take the choice
consistent with the $s$-quark helicity conservation hypothesis, which
is shown by BaBar to be the physical choice~\cite{BaBar-PRD}.
The normalization condition of the amplitudes, 
$
|A_0|^2 + |A_{\parallel}|^2 + |A_{\perp}|^2 = 1
$,
is taken into account by adopting the extended likelihood defined as
$
-\ln L = -\sum_{i=1}^{N_{\rm obs}}\ln {\mathcal G}_i + N_{\rm exp} -
N_{\rm obs}\ln(N_{\rm exp})
$,
where ${\mathcal G}_i$ is the value of the PDF for each event, and
$N_{\rm obs}$ is the number of events used for the fit.
$N_{\rm exp}$ is defined to be $N_{\rm obs} \cdot 
(|A_0|^2+|A_{\parallel}|^2+|A_{\perp}|^2)$ 
to incorporate the normalization condition. 
The normalization of the PDF is recalculated
whenever the fit parameters change. The two charged $B$ decay modes are
combined by defining a single likelihood.

\begin{figure}
\centerline{\mbox{\psfig{figure=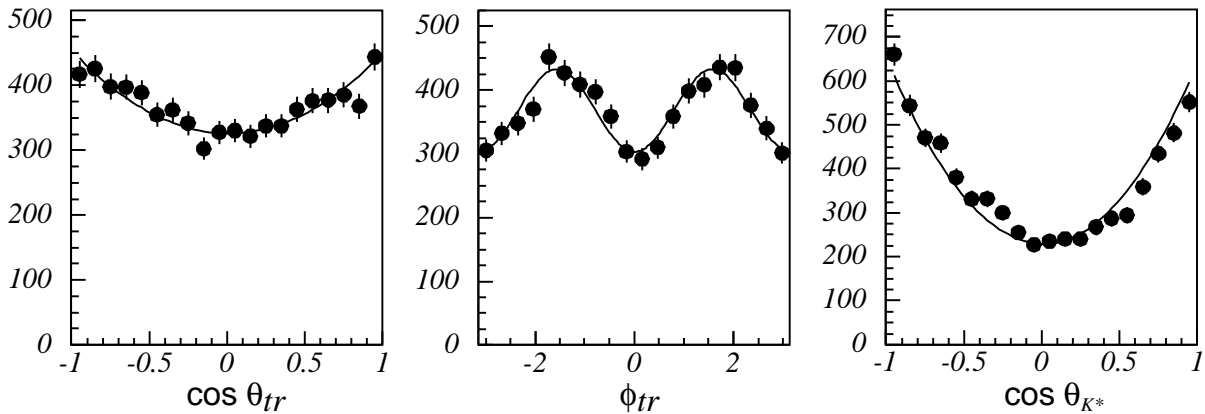,width=16cm}}}
\caption{Distributions of projected angles for $B^0\rightarrow J/\psi
K^{*0}(K^+\pi^-)$. Solid lines show results
of the fit. The data points are corrected for the detector efficiency
and the backgrounds are subtracted.}
\label{fig:angfinal}
\end{figure}

The decay amplitudes determined from the fit are summarized in
Table~\ref{table:amplitude}.
The obtained values are consistent between the two flavors both in neutral 
and charged $B$ decays, indicating no evidence for direct CP violation.
The flavor averaged values 
are consistent with our previous measurement~\cite{angleana-29} 
and that by BaBar using 83 fb$^{-1}$~\cite{BaBar-PRD}. 
Small discrepancies from $\pi$ are observed in
$\arg(A_{\parallel})$ and $\arg(A_{\perp})$ for both $B^0$ and $B^+$
decays. The difference of these two 
phases is $0.458\pm0.110$~rad in $B^0$ decays, 
which is shifted from 0 by
more than $4 \sigma$. This is interpreted as evidence for the existence
of final state interactions. 
Fig.~\ref{fig:angfinal} shows
the projected angular distributions for $B^0\rightarrow J/\psi
K^{*0}(K^+\pi^-)$ decays. 
\begin{table*}
\caption{Measured decay amplitudes for $B^0$ and $B^+$ decays.
The first error is statistical while the second is systematic.}
\begin{center}
\begin{scriptsize}  
\begin{tabular}{c|ccc|ccc}
\hline\hline
   & $B^0$ & $\overline{B}^0$ &  $B^0+\overline{B}^0$ &
   $B^+$ & $B^-$ & $B^+ + B^-$ \\ \hline
$|A_0|^2$  & $0.571\pm0.015$ & $0.578\pm0.016$ &
	     $0.574\pm0.012\pm0.009$ & 
	     $0.600\pm0.020$ & $0.608\pm0.021$ &
             $0.604\pm0.015\pm0.018$  \\

$|A_{\parallel}|^2$ & $0.216\pm0.017$ & $0.244\pm0.018$ &
		      $0.231\pm0.012\pm0.008$ & 
	              $0.194\pm0.019$ & $0.243\pm0.021$ &
		      $0.216\pm0.014\pm0.013$ \\

$|A_{\perp}|^2$  & $0.213\pm0.017$ & $0.178\pm0.017$ &
		   $0.195\pm0.012\pm0.008$ & 
	           $0.206\pm0.019$ & $0.149\pm0.019$ &
                   $0.180\pm0.014\pm0.010$ \\

$\arg(A_{\parallel})$  & $-2.934\pm0.134$ & $-2.851\pm0.114$ &
		         $-2.887\pm0.090\pm0.008$ & 
	                 $-3.070\pm0.142$ & $-3.129\pm0.172$ &
                         $-3.090\pm0.108\pm0.006$ \\

$\arg(A_{\perp})$      & $2.878\pm0.088$ & $2.993\pm0.089$ &
			 $2.938\pm0.064\pm0.010$ & 
	                 $2.964\pm0.099$ & $2.988\pm0.121$ &
                         $2.983\pm0.076\pm0.004$ \\
\hline\hline
\end{tabular}
\label{table:amplitude}
\end{scriptsize} 
\end{center}
\end{table*}

Systematic uncertainties in the fit are determined for:
1) detection efficiency (MC statistics and effect of polarization), 2)
background angular distribution functions, 3) background
fractions, 4) slow pion efficiency, and 5) non-resonant decay 
polarization effect.
The effect of the uncertainty in the fraction of the non-resonant production is
estimated by varying the value by $\pm 5\sigma$ to 
take into account the possible contamination of other resonance 
tails.
The uncertainty in its angular shape is determined by comparing 
with results that assume phase space decay.


The triple product correlations in $B\rightarrow J/\psi K^*$ decays
take the form of $\vec{p}\cdot(\vec{v}_1\times\vec{v}_2)$ in the $B$
rest frame, where
$\vec{p}$ is the momentum of $J/\psi$ or $K^*$, and $\vec{v}_1$($\vec{v}_2$)
is the polarization vector of $J/\psi$($K^*$)~\cite{TP}.
They are odd under time 
reversal (T-odd) and their asymmetries are sensitive to 
direct CP violation even when the strong phase difference is small.
The asymmetries are defined using the decay amplitudes as:
\begin{equation}
A_T^{(1)} = \frac{{\rm Im}(A_{\perp}A_{0}^*)}{A_0^2+A_{\parallel}^2+A_{\perp}^2}, \\
\;\;\;\;
A_T^{(2)} = \frac{{\rm Im}(A_{\perp}A_{\parallel}^*)}{A_0^2+A_{\parallel}^2+A_{\perp}^2}.
\label{eqn:TP}
\end{equation}
The corresponding asymmetries for anti-$B$ decays are
defined as $\bar{A}_T^{(1)}$ and $\bar{A}_T^{(2)}$. The Standard 
Model predicts tiny values for these asymmetries and no difference
between $B$ and anti-$B$ mesons.
Substituting the measured amplitudes in Eq.~\ref{eqn:TP}, the obtained
triple product asymmetries are listed in Table~\ref{table:TP}.
As seen, all the obtained asymmetries are small.
Tiny discrepancies from zero are considered to be
due to final state interactions.
No difference between $A_T^{(1)}$ and $\bar{A}_T^{(1)}$
nor $A_T^{(2)}$ and $\bar{A}_T^{(2)}$ is observed, which is
consistent with the absence of T-odd CP violation.
\begin{table}[h]
\caption{Measured asymmetries in the triple product correlations. 
The first error is statistical while the second is systematic.}
\begin{center}
\begin{tabular}{c|cc}
\hline\hline
 & $B^0$ & $B^+$ \\ \hline
$A_T^{(1)}$        & $0.091 \pm 0.034 \pm 0.007$ & $0.062\pm0.038\pm 0.005$ \\
$A_T^{(2)}$        & $-0.098 \pm 0.032 \pm 0.003$ & $-0.049\pm0.034\pm  0.002$\\
${\bar A}_T^{(1)}$ & $0.047 \pm 0.031 \pm 0.007$ & $0.046\pm0.039\pm 0.005$\\
${\bar A}_T^{(2)}$ & $-0.089 \pm 0.029 \pm 0.003$ & $-0.031\pm0.039\pm 0.002$\\
\hline
$|A_T^{(1)}-{\bar A}_T^{(1)}|$ & $0.044\pm0.046$ & 
	                         $0.016\pm0.054$ \\
$|A_T^{(2)}-{\bar A}_T^{(2)}|$ & $0.009\pm0.043$ &
	                         $0.018\pm0.052$ \\
\hline\hline
\end{tabular}
\label{table:TP}
\end{center}
\end{table}


The CP violation parameters $\sin2\phi_1$ and $\cos2\phi_1$ are 
determined by fitting the time-dependent angular distribution in
Eq.~\ref{eqn:tadf}  to the measured angles and 
$\Delta t$ simultaneously in the CP
decay mode $B^0 \rightarrow J/\psi K^{*0}(K_S^0\pi^0)$ 
using the PDF in Eq.~\ref{eqn:timepdf}.
The procedures to measure $\Delta t$ and to determine
the flavor of the decaying $B^0$ meson are described
elsewhere~\cite{Hadronic}. 
The flavor tagging procedure gives
the flavor $q$ of the tag-side $B$ meson, where $q=+1(-1)$ for
$B^0$($\overline{B}^0$), and the probability $w$ that this 
flavor determination is incorrect.
The value of $\eta$ is given by $-q(1-2w)$ within the signal angular
distribution in Eq.~\ref{eqn:tadf}. The difference in $w$ between
$B^0$ and $\overline{B}^0$ mesons is also considered.

\begin{figure}
\centerline{\mbox{\psfig{figure=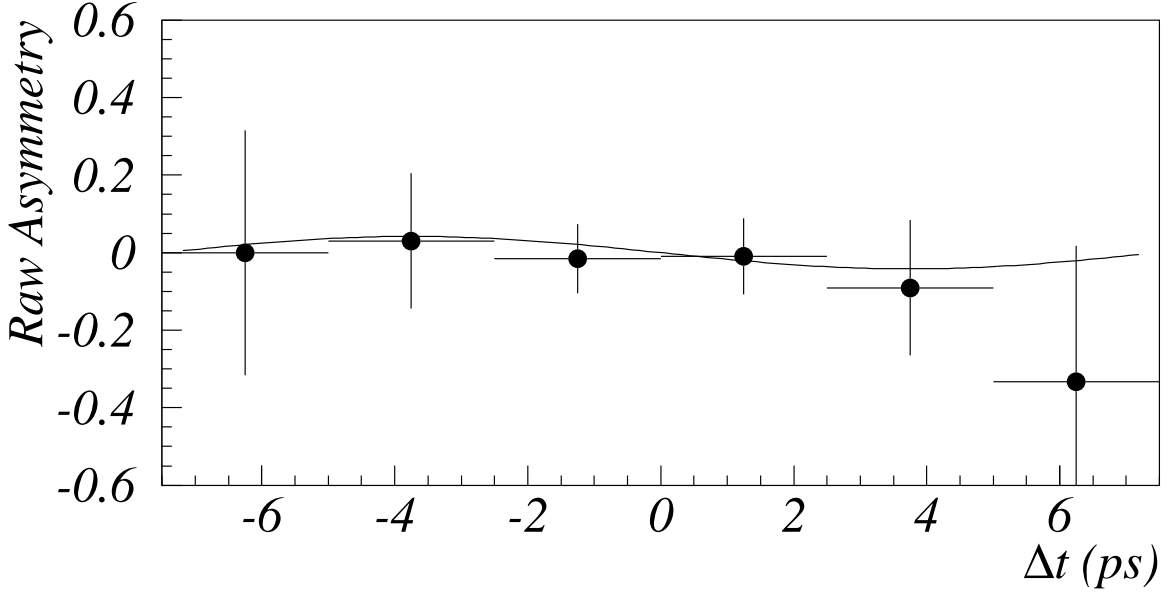,width=16cm}}}
\caption{Raw asymmetry in measured $\Delta t$ between samples tagged
as $q=+1$ and $q=-1$. The solid 
line shows the projection of the fit.}
\label{fig:asym}
\end{figure}

Each term in 
Eq.~\ref{eqn:timepdf}
is convolved with the appropriate resolution
functions separately for the signal, backgrounds having the
$B^0$ lifetime (namely, cross-feeds and non-resonant production),
and the combinatorial background with a zero-lifetime
$\delta$-function shape. The resolution functions are obtained
from fits to the $\Delta t$ distributions measured for various 
data samples. 
Null CP asymmetry is assumed for the backgrounds.
In the fit, the decay amplitudes are fixed at the values obtained
for $B^0$ decays.
The lifetime and mixing parameter are set to PDG values~\cite{PDG}.
From the fit to the data, we obtain
\begin{eqnarray}
	\sin 2\phi_1 & = & 0.24 \pm 0.31 \pm 0.05, \nonumber \\
	\cos 2\phi_1 & = & 0.56 \pm 0.79 \pm 0.11. \nonumber
\end{eqnarray}
When we fix the value of $\sin 2\phi_1$ to the world average value
(0.726)~\cite{HFAG}, the value of $\cos 2\phi_1$ becomes $
0.87 \pm 0.74 \pm 0.12$. The positive sign of $\cos 2\phi_1$ is 
consistent with
the measurement by BaBar~\cite{BaBar-PRD}; however,
we cannot exclude negative values
with our current statistical errors.
The raw asymmetry in the measured $\Delta t$ distribution 
between samples with
$q=+1$ and $-1$ is shown in Fig.~\ref{fig:asym} 
with the projected result of the fit.

Systematic uncertainties in the fit are determined in the same manner as 
those in the $b \to c\bar{c}s$ $\sin 2\phi_1$ measurement~\cite{Hadronic}. 
In addition, the
uncertainties that come from the angular analysis are estimated
similarly as that in the decay amplitude measurement.
The possible bias in the fit is checked by applying the same fitting
procedure to the sample of $B\rightarrow J/\psi K^{*0}(K^+\pi^-)$
decays. We obtain $``\sin 2\phi_1"  =  -0.047 \pm 0.067$ and 
$``\cos 2\phi_1"  =  -0.111 \pm  0.161$, which
are consistent with zero as expected.


In summary, a full angular analysis is performed 
for $B\rightarrow J/\psi K^{*}$ 
decays. The complex decay amplitudes are measured by a
simultaneous fit to three transversity angles.
The measured values are consistent between the  
two $B$ flavors both in 
neutral and charged $B$ meson decays, and 
no direct CP-violating effect is observed.
The difference of arg($A_{\parallel}$) and arg($A_{\perp}$) of $B^0$ decays 
is shifted from 0 by 
more than $4 \sigma$, which is interpreted as evidence for the existence
of final state interactions.
The differences between the asymmetries of triple product correlations 
for $B$ and anti-$B$ mesons are consistent with zero in both 
neutral and charged $B$ mesons,
and no T-odd CP-violating new physics effect is observed. 
The time-dependent angular analysis performed for 
$B^0\rightarrow J/\psi K^{*0}$ ($K^{*0} \rightarrow K_S^0\pi^0$)
decays gives the CP violation parameters
$\sin 2\phi_1 = 0.24 \pm 0.31 \pm 0.05$ and 
$\cos 2\phi_1 = 0.56 \pm 0.79 \pm 0.11$. 
Fixing $\sin 2\phi_1$ at the world average value (0.726) gives 
$\cos 2\phi_1 = 0.87 \pm 0.74 \pm 0.12$. 
The sign of $\cos 2\phi_1$ is positive, although
we cannot exclude negative values with current statistical errors.


We thank the KEKB group for the excellent operation of the
accelerator, the KEK cryogenics group for the efficient
operation of the solenoid, and the KEK computer group and
the NII for valuable computing and Super-SINET network
support.  We acknowledge support from MEXT and JSPS (Japan);
ARC and DEST (Australia); NSFC (contract No.~10175071,
China); DST (India); the BK21 program of MOEHRD and the CHEP
SRC program of KOSEF (Korea); KBN (contract No.~2P03B 01324,
Poland); MIST (Russia); MHEST (Slovenia);  SNSF (Switzerland); NSC and 
MOE(Taiwan); and DOE (USA).


\end{document}